\documentclass{article}
\usepackage{spconf,amsmath,amssymb,graphicx}
\usepackage{bm}
\usepackage{booktabs}
\usepackage{cite}

\DeclareMathOperator*{\argmax}{argmax}
\newcommand{\vb}{\,|\,}
\newcommand{\vB}{\,\Big|\,}
\newcommand{\bx}{\mathbf{x}}
\newcommand{\chk}{\checkmark}

\newcommand{\Oracle}{\textbf{Oracle}}
\newcommand{\BaseA}{\textbf{Base1}}
\newcommand{\BaseB}{\textbf{Base2}}
\newcommand{\Prop}{\textbf{Prop}}

\title{End-to-End Automatic Speech Recognition\\Integrated With CTC-Based Voice Activity Detection}
\name{Takenori Yoshimura$^1$, Tomoki Hayashi$^1$, Kazuya Takeda$^1$, and Shinji Watanabe$^2$}
\address{$^1$Nagoya University, Japan\\$^2$Johns Hopkins University, USA}

\begin{document}
\ninept

\maketitle

\begin{abstract}
This paper integrates a voice activity detection (VAD) function with end-to-end automatic speech recognition toward an online speech interface and transcribing very long audio recordings.
We focus on connectionist temporal classification (CTC) and its extension of CTC/attention architectures.
As opposed to an attention-based architecture, \textit{input-synchronous} label prediction can be performed based on a greedy search with the CTC (pre-)softmax output.
This prediction includes consecutive long blank labels, which can be regarded as a non-speech region.
We use the labels as a cue for detecting speech segments with simple thresholding.
The threshold value is directly related to the length of a non-speech region, which is more intuitive and easier to control than conventional VAD hyperparameters.
Experimental results on unsegmented data show that the proposed method outperformed the baseline methods using the conventional energy-based and neural-network-based VAD methods and achieved an RTF less than $0.2$.
The proposed method is publicly available.\footnote{\url{https://github.com/espnet/espnet}}
\end{abstract}

\section{Introduction}
End-to-end automatic speech recognition (E2E-ASR) has been investigated intensively.
It is a direct mapping from a sequence of acoustic feature vectors into a sequence of graphemes, resulting in eliminating the need for building components requiring expert knowledge in conventional ASR systems, such as morphological analyzers and pronunciation dictionaries.
There are two main types of network architectures for E2E-ASR: connectionist temporal classification (CTC)~\cite{CTC,CTC2} and attention mechanism~\cite{ATT,ATT2}.
A joint framework of CTC- and attention-based architectures~\cite{Hybrid,Hybrid2} has been recently proposed to exploit the advantages of these two architectures and the effectiveness has been shown in various contexts~\cite{RWTH,Suyoun,Stavros,Zhang,Malhotra}.
Although the performance of E2E-ASR improves gradually, most of the approaches rely on the assumption that the input audio recording is appropriately segmented into short audio pieces.
The assumption makes it difficult to use E2E-ASR systems for real-time speech recognition and transcribing unsegmented long audio archives.

There are several attempts to solve the problem.
Modified attention-based methods~\cite{AttentionOnline,AttentionOnline3,Fan,Miao,Niko} restrict the length of the attention window by assuming that the output label sequence is conditioned on a partially observed feature vector sequence rather than the entire feature vector sequence.
Recurrent neural network transducer~(RNN-T)~\cite{RNNT}, which is an extended version of CTC, has also been applied to streaming ASR~\cite{RNNT2,RNNT3} with uni-directional long short-term memory~(LSTM) cells.
Although their effectiveness has been shown, voice activity detection (VAD)~\cite{VAD,VAD2}, which detects the speech segment given an audio sequence, could help to improve the recognition accuracy by discarding input segments that do not contain speech and to reduce the computational cost for processing non-speech segments.
 
The simplest VAD scheme is energy-based: if the short-time energy in a frame of the input audio exceeds a pre-determined threshold, the frame is associated with a voice segment.
Statistical model-based VAD scheme has also been proposed.
In Sohn et al.'s study~\cite{HMM}, hidden Markov models (HMMs) are used to evaluate the likelihood of speech at every frame.
A Gaussian mixture model (GMM) is used as a classifier that distinguishes voice categories from non-voice categories~\cite{GMM2}.
Deep neural network (DNN)-based VAD ~\cite{DBN,DNN,RNN} outperformed conventional VAD schemes recently.
VAD is less dependent on language and does not usually require expert knowledge.
However, building an extra voice activity detector for ASR requires an additional effort.
Furthermore, careful tuning hyperparameters for VAD such as the value of speech/non-speech threshold is often required to maintain a sufficient-level of detection accuracy.
Integrating VAD with ASR in a unified E2E framework is a promising way to eliminate these problems.
There have been some attempts to integrate VAD with conventional HMM-based ASR~\cite{Integration2,Integration3}, but the extensions to E2E-ASR has not been fully investigated.

This paper attempts to integrate VAD with ASR in an E2E manner.
The key idea behind the proposed method is using the CTC architecture.
CTC can perform frame-by-frame label prediction with low computational cost by performing a greedy search.
This input-synchronized label sequence, which cannot be easily obtained in the attention architecture, contains useful information for VAD.
The proposed method uses the number of continuous blank labels as a threshold for detecting speech.
Since CTC does not always activate on the actual timing of phoneme transition in speech~\cite{CTC}, two additional margins are introduced to avoid inappropriate segmentation caused by the behavior.
Although the proposed method is based on heuristic thresholding, the threshold value is directly related to the length of a non-speech region.
Therefore, it is quite intuitive and easy to control compared with hyperparameters in conventional VAD.
Furthermore, the proposed method does not require training of an external voice activity detector, a complex training process to achieve a reasonable performance~\cite{RNNT5}, and high computational cost at inference thanks to the simple strategy.
To support both offline and online speech recognition, we propose two types of inference algorithms using bi-directional and uni-directional architectures.
Note that the proposed method can be integrated with the powerful attention-based encoder-decoder within the hybrid CTC/attention framework~\cite{Hybrid,Hybrid2} and we show the effectiveness of the proposed method based on the framework.

\section{Integrating voice activity detection with automatic speech recognition}
\subsection{CTC-based voice activity detection}
\label{sec:ctc_vad}
E2E-ASR inference is generally defined as a problem to find the most probable grapheme sequence $\hat{C}$ given the audio input $X$:
\begin{equation}
    \hat{C} = \argmax_{C} \, p\left( C \vb X \right),
    \label{eq:bayes_dec}
\end{equation}
where $X = (\bx_1, \ldots, \bx_T)$ is a $T$-length speech feature sequence and $C= (c_1, c_2, \ldots)$ is a sequence of grapheme symbols such as alphabetical letters and Kanji characters.
To handle very long audio recordings, we assume that $\hat{C}$ consists of the most probable partial grapheme sequences $\{ \hat{C}^{(i)} \}_{i=1}^N$ given the non-overlapping partial audio subsequences $\{ X^{(i)} \}_{i=1}^N$:
\begin{equation}
    \hat{C} \simeq \argmax_{C} \, \prod_{i=1}^N p\left( C^{(i)} \vB X^{(i)} \right),
\end{equation}
where $N$ is the number of speech segments.
The problem of VAD is how to segment $X$ into $X^{(1)}, \, \ldots, \, X^{(N)}$. 

In this paper, the label sequence obtained by performing frame-by-frame prediction based on a greedy search with the CTC softmax output is used as a cue to solve the problem.
The CTC formulation is started from the posterior probability $p\left( C \vb X \right)$ introduced in Eq.~\eqref{eq:bayes_dec}, which is factorized based on the conditional independence assumption as follows:
\begin{equation}
    p\left( C \vb X \right) = \!\! \sum_{Z \in \mathcal{Z}(C)} \!\! p\left(Z \vb X\right) \, \approx \!\! \sum_{Z \in \mathcal{Z}(C)} \prod_{t \in \mathcal{T}} p\left(z_t \vb X\right),
\label{eq:ctc}
\end{equation}
where $Z$ is a $K$-length CTC state sequence composed of the original grapheme set and the additional blank label, $\mathcal{Z}(C)$ is a set of all possible sequences given the grapheme sequence $C$, and $\mathcal{T}=\{ r, 2r, 3r, \ldots \}$ is a set of subsampled time indices with the subsampling factor $r$ $(K \times r \simeq T)$.
We focus on the frame-level posterior distribution $p\left(z_t \vb X\right)$ in Eq.~\eqref{eq:ctc} because it provides \textit{frame-synchronous information} obtained in the early stage of the inference process.
With the greedy search algorithm, the following frame-synchronous label output can be obtained without performing the time-consuming beam search algorithm:
\begin{align}
    \hat{z}_t = \argmax_{z_t} \, p\left(z_t \vb X\right).
\end{align}
This operation corresponds to simply taking the label with the highest probability in the categorical distribution of the CTC softmax output at each time step.\footnote{The output before applying softmax function, which is called the pre-softmax output, can be used instead for saving computational costs.}
The predicted label $\hat{z}_t$ is allowed to be the blank label that often represents non-speech segments in addition to the repetition of the grapheme symbol, i.e., the output label sequence of CTC is expected to contain helpful information for VAD.

\begin{figure}[t]
  \begin{center}
    \includegraphics[width=0.99\columnwidth]{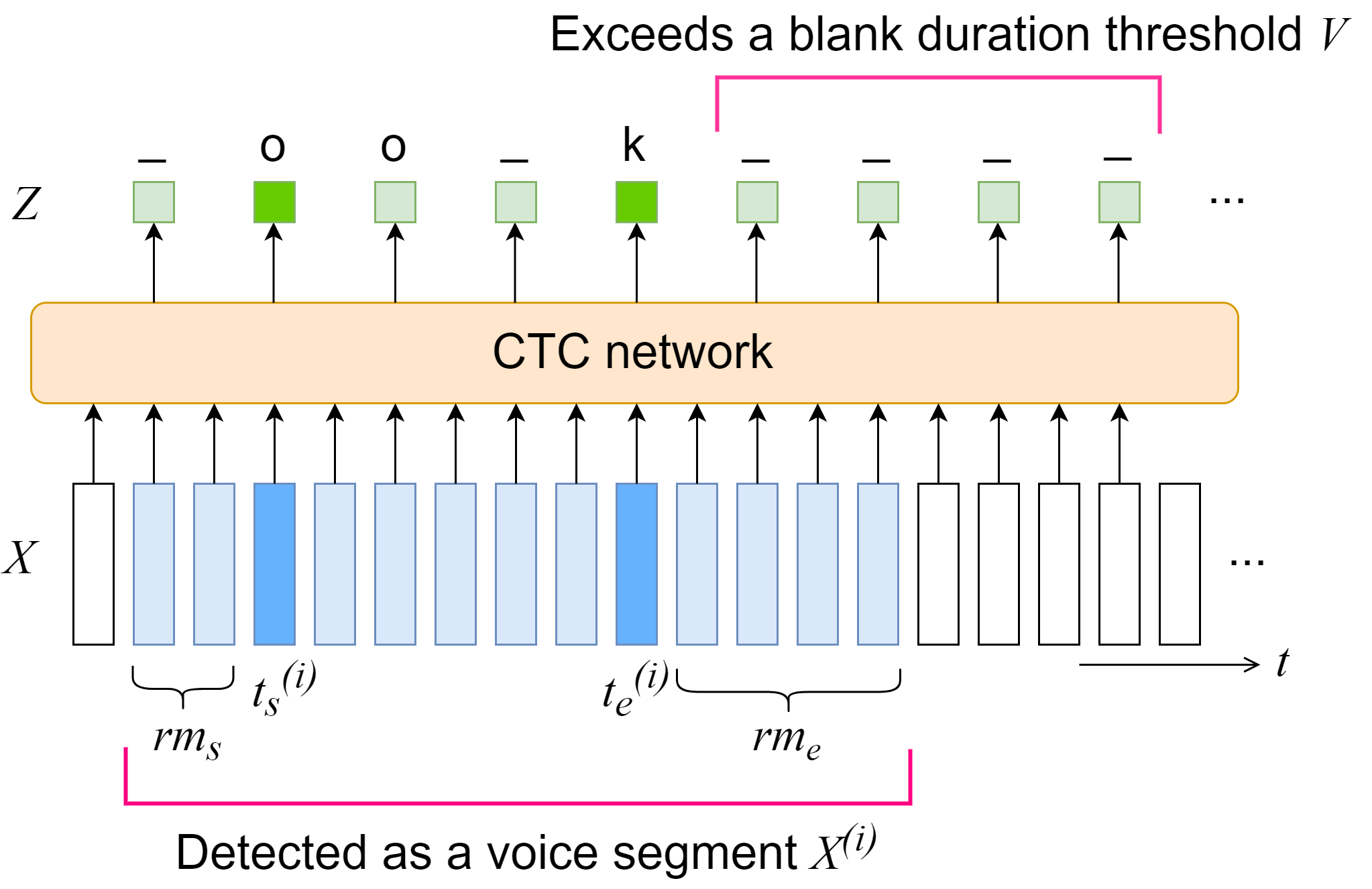}
    \vspace{-1mm}
    \caption{Example of detected voice segments in the proposed method where subsampling factor $r=2$, minimum blank duration threshold $V=4$, onset margin $m_s=1$, and offset margin $m_e=2$. The shaded acoustic feature vectors will be used for predicting a final label sequence.}
    \label{fig:margin}
    \vspace{-4mm}
  \end{center}
\end{figure}

Figure~\ref{fig:margin} shows an overview of the proposed method. As shown in the figure, if the number of consecutive blank labels exceeds a pre-determined threshold $V$ (which is referred as to \textit{minimum blank duration threshold}) given a sequence of $\hat{z}_t$, the input audio is segmented based on the position of non-blank labels:
\begin{equation}
    X^{(i)} = \left( \begin{array}{cccc} \bx_{t_s^{(i)}}, & \bx_{t_{s}^{(i)}+1}, & \ldots, & \bx_{t_e^{(i)}} \end{array} \right),
\end{equation}
where $t_s^{(i)}$ denotes an index corresponding to the first non-blank label after the $i-1^\mathrm{th}$ segment, and $t_e^{(i)}$ denotes an index corresponding to the last non-blank label before consecutive blank labels. Since there is no guarantee that the positions of non-blank labels $t_s^{(i)}$ and $t_e^{(i)}$ correspond to the actual beginning/end of phonemes due to the properties of CTC, we introduce two margins as
\begin{equation}
    X^{(i)} = \left( \begin{array}{cccc} \bx_{t_s^{(i)} - rm_s}, & \ldots, & \bx_{t_e^{(i)} + rm_e} \end{array} \right),
\end{equation}
where $m_s$ and $m_e$ are referred to as \textit{onset margin} and \textit{offset margin}, respectively, and $r$ is the subsampling factor as described before.

\subsection{Offline and online inference}
Bi-directional LSTM cells rather than uni-directional ones are typically used in the CTC encoder.
The inference algorithm of the proposed method using the bi-directional encoder consists of two stages:
in the first stage, a mixed blank and non-blank label sequence is obtained by the greedy search given the entire audio input sequence, and in the second stage, a final label sequence is obtained by feeding the partial audio input sequences extracted by using the mixed label sequence.
This inference algorithm is suitable for offline recognition, e.g., transcribing long audio recordings, because of the inevitable delay caused by the two-path decoding.
On the other hand, online decoding can be performed by using the uni-directional encoder because there is no need to recompute the encoder states, which differs from the first algorithm.
The algorithmic delay mainly depends on the length of the offset margin.
This seems to be reasonable for practical use.

Although the proposed method depends on CTC, it can be combined with the attention decoder within the hybrid CTC/attention framework~\cite{Hybrid}.
In this framework, the encoders for both CTC and attention models are shared and trained based on their two objective functions.
The decoders for both CTC and attention models can be used for decoding simultaneously, resulting in better recognition performance.

\section{Experiments}
Two datasets were used to investigate the effectiveness of the proposed method.
Note that the datasets consist of spontaneous talks that sometimes exceed 20 minutes in length. Audio segmentation is an essential technique for saving computational resources and complexity.
ESPnet~\cite{ESP} was used as an E2E-ASR toolkit through the experiments.
The following four methods were compared:
\begin{itemize}
\setlength{\itemsep}{0mm}
\item \Oracle: The audio input of the E2E model was segmented according to the time information provided by the datasets.
\item \BaseA: The audio input of the E2E model was automatically segmented by the simple energy-based VAD algorithm, which is implemented on the Kaldi toolkit.\footnote{\url{https://github.com/kaldi-asr/kaldi}}
The hyperparameters of the algorithm such as the energy threshold were tuned by a grid search.
\item \BaseB: The audio input of the E2E model was automatically segmented by data-driven neural-network-based VAD.
We built a robust voice activity detector across different corpora trained by large amounts of speech segments with noisy reverberant data augmentation.
% the neural-network was trained on the external data from 100 hours of the LibriSpeech corpus~\cite{librispeech}.
% The training data was perturbed with various room impulse responses, additive noise, and music from the MUSAN corpus~\cite{musan}.
The network consisted of five time-delay neural network layers and two statistics pooling layers.
The number of hidden units per layer was 256.
The input of the network was 40-dimensional MFCC with around $800$ ms of the left context and $200$ ms of the right context.
Simple Viterbi decoding on an HMM is used to obtain speech activity.
\item \Prop: The audio input of the E2E model was automatically segmented based on the proposed method described in the previous section.
\end{itemize}
The E2E model was mainly based on the hybrid CTC/attention architecture and trained on the segmented data provided by the datasets.

\subsection{Japanese dataset}
The CSJ corpus~\cite{CSJ} was used for our first evaluation.
It contains about 650 hours of spontaneous Japanese speech data.
The evaluation data is composed of three sets: \textsl{eval1}, \textsl{eval2}, and \textsl{eval3}.
The training hyperparameters were set to the default values of the training recipe provided by ESPnet except that the mini-batch size and the number of units were half the default value.
Assuming offline recognition, the network structure consisted of a 4-layer bi-directional LSTM encoder and a 1-layer LSTM decoder.
The subsampling factor $r$ introduced in Section~\ref{sec:ctc_vad} was set to four.
The input feature was composed of 80-dimensional mel-scale filter-bank features and 3-dimensional pitch features, which were calculated every 10 ms.
The output of the E2E model included Japanese Kanji and Hiragana characters.
To avoid obtaining very short recognition results against input audio length due to the attention decoder predicting the early appearance of the end-of-sentence symbol, the recognition candidates satisfying the following equation were rejected:
\begin{equation}
    \frac{\mbox{Length of output label sequence}}{\mbox{Length of subsampled encoded sequence}} \le \alpha,
    \label{eq:minlen_ratio}
\end{equation}
where $\alpha$ was set to $0.1$ in the experiments.
This scheme was very important to avoid the effect of the unintended behavior of the attention mechanism.
Character error rate~(CER) was used as the evaluation metric because CER is widely used for the Japanese ASR evaluation due to its ambiguous word boundary.

\subsubsection{Decoding hyperparameters}
The initial experiment only used the CTC decoding instead of the joint CTC/attention decoding.
We first investigated the effect of the value of the onset margin $m_s$, as introduced in Section~\ref{sec:ctc_vad}.
It was varied from $0$ to $3$ while minimum blank duration threshold $V=16$ and offset margin $m_e=2$.
Table~\ref{tab:csj_onset} summarizes the CERs obtained in the experiment.
It can be said from the table that the onset margin is important to achieve correct recognition results.
The value of $m_e$ was also varied from $0$ to $3$ while $V=16$ and $m_s=2$.
The result is shown in Table~\ref{tab:csj_offset}.
The performance difference between with and without the offset margin, $m_e > 0$ and $m_e=0$, was significantly large.
These two margins seem to buffer the time-lag of the CTC softmax output.

\begin{table}[t]
  \begin{center}
    \caption{Effect of value of onset margin.}
    {\renewcommand\arraystretch{0.9}
    %\vspace{-2mm}
    \begin{tabular}{ccccc} \toprule
      $m_s$        & eval1 [\%] & eval2 [\%] & eval3 [\%] & Avg. [\%] \\ \midrule
      0 & 10.4 & 7.2 & 8.6 & 8.7 \\
      1 & 10.1 & 7.0 & 8.1 & 8.4 \\
      2 & 10.1 & 7.0 & 8.0 & \textbf{8.4} \\
      3 & 10.1 & 7.1 & 8.1 & 8.4 \\ \bottomrule 
    \end{tabular}
    \label{tab:csj_onset}}
    %\vspace{-4mm}
  %\end{center}
%\end{table}
%\begin{table}[t]
  %\begin{center}
    \caption{Effect of value of offset margin.}
    {\renewcommand\arraystretch{0.9}
    %\vspace{-2mm}
    \begin{tabular}{ccccc} \toprule
      $m_e$    & eval1 [\%] & eval2 [\%] & eval3 [\%] & Avg. [\%] \\ \midrule
      0 & 10.7 & 7.5 & 8.9 & 9.0 \\
      1 & 10.2 & 7.1 & 8.3 & 8.5 \\
      2 & 10.1 & 7.0 & 8.0 & 8.4 \\
      3 & 10.0 & 6.9 & 7.9 & \textbf{8.3} \\ \bottomrule
    \end{tabular}
    \label{tab:csj_offset}}
    %\vspace{-4mm}
  %\end{center}
%\end{table}
%\begin{table}[t]
  %\begin{center}
    \caption{Effect of value of minimum blank duration threshold.}
    {\renewcommand\arraystretch{0.9}
    %\vspace{-2mm}
    \begin{tabular}{ccccc} \toprule
      $V$        & eval1 [\%] & eval2 [\%] & eval3 [\%] & Avg. [\%] \\ \midrule
       8 & 11.2 & 8.2 & 8.9 & 9.7 \\
      12 & 10.5 & 7.3 & 8.3 & 8.7 \\
      16 & 10.1 & 7.0 & 8.0 & 8.4 \\
      20 & 10.0 & 6.8 & 7.7 & 8.2 \\
      24 &  9.9 & 6.7 & 6.7 & \textbf{7.8} \\ \bottomrule
    \end{tabular}
    \label{tab:csj_threshold}}
    \vspace{-3mm}
  \end{center}
\end{table}

To investigate the sensitivity to the change of the minimum blank duration threshold, the value of $V$ was varied with the fixed onset and offset margins ($m_s=2$ and $m_e=2$).
Table~\ref{tab:csj_threshold} shows the results against all the test sets.
From the table, a low CER can be obtained by selecting a large threshold value, as expected.
However, reasonable recognition performance could be achieved by a medium threshold value, e.g., $V=16$.
This means that $640~(= 16 \times 4 \times 10)$ ms was used as a threshold value.
The duration seems to be reasonable for detecting short pause, indicating that the proposed method can intuitively tune VAD parameters in contrast to the conventional VAD methods requiring non-intuitive parameters such as an energy threshold.

\subsubsection{Performance comparison}
In the next experiment, we compared the baseline methods with the proposed one whose hyperparameters were set to $V=16$, $m_s=2$, and $m_e=3$.
Table~\ref{tab:csj_comp} shows the results of the experiments using not only the CTC decoder but also the attention decoder and the joint CTC/attention decoder.
\BaseB{} obtained a lower CER than \BaseA.
This indicates the effectiveness of the data-driven VAD method.
\Prop{} achieved further improvement over \BaseB.
CTC is a mapping from acoustic features to characters whereas the standard data-driven VAD is a mapping from acoustic features to a simple binary symbol indicating speech or non-speech.
The performance gain may be because the proposed method was able to detect speech considering linguistic information provided by CTC.
\Prop{} obtained the best CER regardless of the decoder type.
In the following experiments, the joint decoding using the CTC prefix score was used.

\subsection{English dataset}
The TED-LIUMv2 corpus~\cite{TED}, which is a set of English TED talks with transcriptions, was used for our second evaluation.
It contains about 200 hours of speech data.
The training hyperparameters were set to the default values of the training recipe provided by ESPnet.
Two types of the network structure were investigated: 6-layer bi-directional LSTM encoder with 1-layer LSTM decoder, and 6-layer uni-directional LSTM encoder with 2-layer LSTM decoder.
The output of the E2E model was subword units encoded by byte-pair encoding~(BPE) instead of alphabetical letters.
A 4-layer LSTM-based language model was trained using the same corpus and was used at the decoding stage.
Word error rate~(WER) as well as CER were used as the evaluation metric.

\subsubsection{Decoding hyperparameters}
In the first experiment, the effects of the onset and offset margins were investigated.
The minimum blank duration threshold was fixed as $V=16$, which comes from the result of the previous evaluation.
Tables~\ref{tab:ted_onoffset_bi} and \ref{tab:ted_onoffset_uni} show the experimental results of the bi-directional and uni-directional cases, respectively.
For the bi-directional case, the large offset margin was important whereas the large onset margin was important for the uni-directional case.
The appropriate values of the two margins differed from those of the previous evaluation.
This may be because of the difference in the type of subword units.

\begin{table}[t]
  \begin{center}
    \caption{Performance comparison on CSJ corpus.}
    {\renewcommand\arraystretch{0.9}
    %\vspace{-2mm}
    \begin{tabular}{cccc} \toprule
                & CTC [\%] & Attention [\%] & Joint [\%] \\ \midrule
      \Oracle   & 7.2 & 7.7 & 6.1 \\ \midrule
      \BaseA    & 8.9 & 9.4  & 8.6 \\
      \BaseB    & 8.6 & 8.9  & 8.1 \\
      \Prop     & \textbf{9.2} & \textbf{8.3} & \textbf{7.6} \\ \bottomrule 
    \end{tabular}
    \label{tab:csj_comp}}
    %\vspace{-4mm}
  %\end{center}
%\end{table}
%\begin{table}[t]
  %\begin{center}
    \caption{Effect of decoding hyperparameters (bi-directional).}
    %\vspace{1mm}
    {\renewcommand\arraystretch{0.9}
    \begin{tabular}{cccc} \toprule
      $m_s$ & $m_e$ & dev [\%] & test [\%] \\ \midrule
      4 & 2  & 15.2 & 19.5 \\
      4 & 6  & 13.2 & 16.6 \\
      4 & 10 & \textbf{12.2} & \textbf{15.4} \\
      2 & 10 & 12.3 & 15.6 \\
      \bottomrule
    \end{tabular}
    \label{tab:ted_onoffset_bi}}
    %\vspace{-2mm}
  %\end{center}
%\end{table}
%\begin{table}[t]
  %\begin{center}
    \caption{Effect of decoding hyperparameters (uni-directional).}
    %\vspace{1mm}
    {\renewcommand\arraystretch{0.9}
    \begin{tabular}{cccc} \toprule
      $m_s$ & $m_e$ & dev [\%] & test [\%] \\ \midrule
       2 & 4 & 20.9 & 27.0 \\
       6 & 4 & 16.0 & 20.4 \\
      10 & 4 & 14.9 & 19.0 \\
      10 & 2 & \textbf{14.9} & \textbf{18.9} \\
      \bottomrule
    \end{tabular}
    \label{tab:ted_onoffset_uni}}
    \vspace{-3mm}
  \end{center}
\end{table}

\subsubsection{Performance comparison}
In the next experiment, the proposed method was compared with the baseline methods where the hyperparameters of \Prop{} were set to $V=16$, $m_s=4$, and $m_e=10$ for the bi-directional case and $V=16$, $m_s=10$, and $m_e=2$ for the uni-directional case.
Tables~\ref{tab:ted_bi} and \ref{tab:ted_uni} summarize the obtained CERs and WERs.
The proposed method again outperformed the baseline methods for both the cases, indicating the effectiveness of the proposed method.

\subsubsection{Inference speed}
In the final experiment, real-time factors~(RTFs) of the proposed method with the uni-directional encoder were measured under some conditions.
To speed up inference, the vectorized beam search algorithm~\cite{Vector} was used. The margin parameter in the algorithm was set to 50.
CPU with four threads or a single-GPU (NVIDIA TITAN~V) was used.
The obtained RTFs as well as WERs are shown in Table~\ref{tab:ted_rtf}.
The vectorized beam search improved RTF with small accuracy degradation.
As expected, GPU accelerated the inference speed significantly and an RTF of $0.18$ was achieved.
This indicates that the proposed method is acceptable for real-time applications.

\section{Conclusion}
Automatic speech recognition and voice activity detection were integrated in an end-to-end manner.
Our experiments showed the effectiveness of the proposed method over conventional voice activity detectors.
Future work includes applying the proposed idea to the Transformer, which is a promising model for outperforming the LSTM encoder-decoder model~\cite{Karita}.

\section{Acknowledgements}
This research is supported by the Center of Innovation~(COI) program from the Japan Science and Technology Agency~(JST).
The authors thank Vimal Manohar for kindly providing us with a DNN-based voice activity detector.

\begin{table}[t]
  \begin{center}
    \caption{Performance comparison on TED-LIUM corpus (bi-directional).}
    \vspace{0.5mm}
    {\renewcommand\arraystretch{0.9}
    \begin{tabular}{ccc} \toprule
                & dev CER~/~WER [\%] & test CER~/~WER [\%] \\ \midrule
      \Oracle   &  9.6~/~11.7 &  9.4~/~11.2 \\ \midrule
      \BaseA    & 11.2~/~13.0 & 16.6~/~17.6 \\
      \BaseB    & 10.7~/~12.9 & 14.9~/~17.2 \\
      \Prop     & \textbf{10.0}~/~\textbf{12.2} &
                  \textbf{13.3}~/~\textbf{15.4} \\ \bottomrule 
    \end{tabular}
    \label{tab:ted_bi}}
    %\vspace{-2mm}
  %\end{center}
%\end{table}
%\begin{table}[t]
  %\begin{center}
    \caption{Performance comparison on TED-LIUM corpus (uni-directional).}
    \vspace{0.5mm}
    {\renewcommand\arraystretch{0.9}
    \begin{tabular}{ccc} \toprule
                & dev CER~/~WER [\%] & test CER~/~WER [\%] \\ \midrule
      \Oracle   & 11.9~/~14.1 & 13.0~/~14.9 \\ \midrule
      \BaseA    & 14.0~/~15.7 & 20.2~/~20.3 \\
      \BaseB    & 13.4~/~15.9 & 18.8~/~20.7 \\
      \Prop     & \textbf{12.5}~/~\textbf{14.9} &
                  \textbf{16.8}~/~\textbf{18.9} \\ \bottomrule 
    \end{tabular}
    \label{tab:ted_uni}}
    %\vspace{-2mm}
  %\end{center}
%\end{table}
%\begin{table}[t]
  %\begin{center}
    \caption{Speed improvement of online inference.}
    %\vspace{1mm}
    {\renewcommand\arraystretch{0.9}
    \begin{tabular}{ccccc} \toprule
      Vectorization & GPU & dev [\%] & test [\%] & RTF \\ \midrule
           &      & 14.9 & 18.9 & 3.77 \\
      \chk &      & 15.0 & 19.0 & 2.93  \\
           & \chk & 14.9 & 18.9 & 0.27 \\
      \chk & \chk & 15.0 & 19.0 & \textbf{0.18} \\
      \bottomrule 
    \end{tabular}
    \label{tab:ted_rtf}}
    \vspace{-3mm}
  \end{center}
\end{table}

\bibliographystyle{IEEEbib}
\bibliography{mybib}

\end{document}